\documentclass[cpp,a4paper,fleqn%
]{w-art}
\usepackage{times,cite,w-thm,amsmath,amssymb}
\theoremstyle{plain}

\theoremstyle{definition}

\usepackage[]{graphicx}
\usepackage{color}

\newcommand{\vc}{\mathbf}

\newcommand{\B}{\textrm{B}}
\newcommand{\arctanh}{\textrm{arctanh}}

\newcommand{\ident}{\ensuremath{\mathcal{I}}}
\newcommand{\erf}{\ensuremath{\textrm{erf}}}

\begin{document}
\DOIsuffix{theDOIsuffix}
\Volume{46}
\Month{01}
\Year{2007}
\pagespan{1}{}
\Receiveddate{XXXX}
\Reviseddate{XXXX}
\Accepteddate{XXXX}
\Dateposted{XXXX}



\title[Temperature Anisotropy Relaxation of the OCP]{Temperature Anisotropy Relaxation of the One-Component Plasma}


\author[S.D.\ Baalrud]{Scott D.\ Baalrud\inst{1,}%
  \footnote{Corresponding author\quad E-mail:~\textsf{scott-baalrud@uiowa.edu},
            Phone: +01\,319\,335\,1695,
            Fax: +01\,319\,335\,1753}}
\address[\inst{1}]{Department of Physics and Astronomy, University of Iowa, Iowa City, Iowa 52242, USA}
\author[J.\ Daligault]{J\'er\^ome Daligault  \inst{2}} 
\address[\inst{2}]{Theoretical Division, Los Alamos National Laboratory, Los Alamos, NM 87545, USA}
\begin{abstract}

The relaxation rate of a Maxwellian velocity distribution function that has an initially anisotropic temperature $(T_\parallel \neq T_\perp)$ is an important physical process in space and laboratory plasmas. It is also a canonical example of an energy transport process that can be used to test theory.  Here, this rate is evaluated using molecular dynamics simulations of the one-component plasma. Results are compared with the predictions of four kinetic theories; two treating the weakly coupled regime (1) the Landau equation, and (2) the Lenard-Balescu equation, and two that attempt to extend the theory into the strongly coupled regime (3) the effective potential theory and (4) the generalized Lenard-Balescu theory. The role of dynamic screening is studied, and is found to have a negligible influence on this transport rate. Oscillations and a delayed relaxation onset in the temperature profiles are observed at strong coupling, which are not described by the kinetic theories. 
 
\end{abstract}
\maketitle                   






\section{Introduction\label{sec:intro}}

The collisional relaxation of a temperature anisotropy is a canonical example of energy transport in plasmas~\cite{ichi:70}. It serves as a test for transport theories, as well as playing an essential role in many experiments~\cite{hyat:87,beck:92,dubi:05,ande:09} and natural occurring plasmas~\cite{maru:11}. For instance, plasmas that are preferentially heated, or cooled, in one direction will form a temperature anisotropy. Magnetized plasmas often have different energy confinement times either along or against the magnetic field, and can form a temperature anisotropy as a result~\cite{ott:17}. It is often essential to know the rate at which this temperature anisotropy relaxes in order to accurately model these plasmas. Here, we use molecular dynamics (MD) simulations to test theories of the collisional temperature anisotropy relaxation rate. The work concentrates on the unmagnetized one-component plasma (OCP)~\cite{baus:80} and considers a broad range of coupling parameters. Although this concentrates on unmagnetized plasmas, the results are also expected to apply to magnetized plasmas as long as $\omega_c/\omega_p \ll 1$, where $\omega_c = eB/m$ is the gyrofrequency and $\omega_p = \sqrt{4\pi e^2n/m}$ is the plasma frequency~\cite{ichi:70,dong:13}. Examples of systems in which moderate to strong coupling arises and magnetization may lead to temperature anisotropies include magnetized ultracold neutral plasmas~\cite{zhan:08} and in magnetized inertial confinement fusion~\cite{gome:14}.

The initial distribution function is assumed to have the form 
\begin{equation}
f = \frac{n}{\pi^{3/2} v_{T\parallel} v_{T\perp}^2} e^{-v_\parallel^2/v_{T\parallel}^2} e^{-v_\perp^2/v_{T\perp}^2}  \label{eq:amax}
\end{equation}
where $n$ is the number density, $v_{T\parallel}^2 = 2k_BT_\parallel/m$ and $v_{T\perp}^2 = 2k_BT_\perp/m$. Here, $T_\parallel$ and $T_\perp$ are the parallel and perpendicular temperatures, which are related to the total temperature by $T = (T_\parallel + 2 T_\perp)/3$. In weak and moderately coupled plasmas, the rate of relaxation of the temperature anisotropy can be modeled from the $v_\parallel^2$ or $v_\perp^2$ moment of a kinetic equation $df/dt = C(f)$. This provides the anisotropy evolution equation 
\begin{equation}
\frac{dT_\perp}{dt} = - \frac{1}{2} \frac{d T_\parallel}{dt} = - \nu (T_\perp - T_\parallel) \label{eq:nrl}
\end{equation}
where  
\begin{equation}
\nu = \frac{m}{n} \int d^3v\, v_\perp^2 \frac{C(f)}{T_\perp - T_\parallel} = - \frac{1}{2} \frac{m}{n} \int d^3v\, v_\parallel^2 \frac{C(f)}{T_\perp - T_\parallel} \label{eq:nu_def}
\end{equation}
is the collisional equipartition rate and $C(f)$ is the collision operator. 

A consequence of the underlying assumptions of kinetic theories of weak or moderately coupled plasmas is that kinetic energy is a conserved quantity throughout the relaxation $dT/dt=0$.  However, this assumption becomes suspect at strong coupling where the potential energy of interactions in the system is high enough that the exchange between kinetic and potential energy may significantly influence  the relaxation processes. The MD simulation results shown here reveal that, indeed, kinetic energy is well conserved at weak and moderate coupling, but that small dynamic oscillations in the kinetic energy, due to exchange with potential energy, are observed at strong coupling. Recent MD simulations have also shown that external energy perturbations can lead to the spontaneous development of temperature anisotropies~\cite{ott:17}. 

Each of the theories that will be discussed assume kinetic energy conservation, and thus satisfy Eq.~(\ref{eq:nrl}). This is a set of two coupled ordinary differential equations that can be solved for $T_\parallel(t)$ and $T_\perp(t)$. For comparing different collision theories, it is convenient to note that the equations can be decoupled to a single equation by defining  
\begin{equation}
\label{eq:Adef}
A \equiv \frac{T_\perp}{T_\parallel} - 1
\end{equation}
as a measure of the temperature anisotropy. Note that $T>0$ implies $A>-1$. In terms of $A$, Eq.~(\ref{eq:nrl}) is 
\begin{equation}
\frac{dA}{dt} = - \nu (3A + 2 A^2) . \label{eq:dAdt}
\end{equation} 
Thus, $A(t)$ along with $dT/dt = 0$ provides the solution for both the parallel $T_\parallel(t) = 3T/(3+2A(t))$ and perpendicular $T_\perp(t) = 3T (1+A(t))/(3+2A(t))$ temperatures. 

Here, we compute $\nu$ from four different collision operators and compare the results with MD simulations. Simulations were carried out spanning a wide range of coupling strengths. The coupling strength of the OCP is described by the Coulomb coupling parameter~\cite{baus:80} 
\begin{equation}
\Gamma \equiv \frac{e^2/a}{k_BT} ,
\end{equation}
where $e$ is the elementary electron charge,  $T$ is the total temperature and $a=(3/4\pi n)^{1/3}$ is the average interparticle spacing. Two of the collision operators we compare with apply to weakly coupled plasmas ($\Gamma \ll 1$): the Landau equation~\cite{land:36} and the Lenard-Balescu equation~\cite{lena:60}. The distinguishing feature of these is that the Lenard-Balescu equation treats dynamic (velocity-dependent) screening, whereas the Landau equation is based on a static screening model. We find that both give very similar predictions for $\Gamma \lesssim 0.1$ and that both models rapidly break down for larger $\Gamma$ values. Thus, there is no significant contribution associated with dynamic screening for this process over the range of coupling strength in which these theories are applicable. Physically, this is because temperature anisotropy relaxation is a thermal processes associated with particles in the thermal part of the distribution, which do not have significant velocity-dependent components to their screening clouds. We also compare with two theories that extend these collision operators to strong coupling: effective potential theory (EPT)~\cite{baal:13,baal:14,baal:15} and generalized Lenard-Balescu theory (GLB)~\cite{tana:86,ichi:92,bene:12}. Again, GLB includes dynamic effects in the long distance component of the linear response function, whereas EPT is based on a static description. We find that each of these give similar results that are in agreement with MD from weak to moderate coupling, but that EPT is able to extend to significantly larger $\Gamma$ values than the GLB theory. Again, the conclusion is that dynamic screening does not significantly influence the anisotropy relaxation rate. 

The main approximation in this analysis is that the distribution function maintains the anisotropic Maxwellian form from Eq.~(\ref{eq:amax}) throughout the relaxation, with only the magnitudes of $T_\parallel$ and $T_\perp$ evolving. It is known from other transport processes, such as self-diffusion of the OCP, that deviations from Maxwellian distributions can contribute corrections on the order of $20\%$ in the weakly coupled limit, but are typically insignificant in the strongly coupled limit~\cite{ferz:72,baal:14}. To the best of our knowledge, a perturbation expansion of the distribution function, such as the Chapman-Enskog~\cite{ferz:72,chap:70} or Grad~\cite{grad:58} expansions for hydrodynamic transport coefficients, has not yet been developed for temperature anisotropy relaxation. This aspect of the analysis will be discussed further in Sec.~\ref{sec:results}. 

In addition to comparing with MD simulations, explicit expressions are derived for $\nu$ from each collision operator. Various limits of each are also considered, including early time, weak anisotropy and limits connecting the theories based on static or dynamic screening. The results quantify the accuracy of each approach, and the various reduced models in asymptotic limits. This provides useful information for modeling a variety of different plasmas with temperature anisotropies.  

Finally, time-dependent temperature profiles from the MD simulations are used to highlight interesting features of correlation effects that arise at strong coupling, which are not described by any of the theories considered. These include delays in the temperature relaxation at early times, substantial oscillations in the temperature profiles in time, and small violations of the kinetic energy conservation assumption ($dT/dt=0$), indicating correlation between kinetic and internal energies during the relaxation process. These aspects are discussed in Sec.~\ref{sec:results}.

\section{Weakly Coupled Plasma Theories\label{sec:wc}}

\subsection{Landau theory}

Plasma transport theory is typically derived from the Landau kinetic equation~\cite{land:36}, which for the OCP is
\begin{equation}
C_\textrm{L} = - \frac{2\pi e^4 \ln \Lambda}{m^2} \frac{\partial}{\partial \vc{v}} \cdot \int d^3v^\prime \frac{u^2 \ident - \vc{u} \vc{u}}{u^3} \cdot \biggl[ f(\vc{v}) \frac{\partial f(\vc{v}^\prime)}{\partial \vc{v}^\prime} - f(\vc{v}^\prime) \frac{\partial f(\vc{v})}{\partial \vc{v}} \biggr] \label{eq:c_landau}
\end{equation}
where $\vc{u} \equiv \vc{v} - \vc{v}^\prime$ and $\ln \Lambda$ is the Coulomb logarithm. In Landau's theory $\Lambda = b_\textrm{max}/b_\textrm{min}=\lambda_D/r_L$ is the ratio of the maximum and minimum length scales over which particles interact. The maximum is determined by the Debye screening cloud $\lambda_D = \sqrt{k_BT/(4\pi e^2n)}$ and the minimum by the distance of closest approach $r_L = e^2/(2k_BT)$.  In terms of the Coulomb coupling parameter $\Lambda = 2/(\sqrt{3} \Gamma^{3/2})$. The standard plasma physics result for the temperature anisotropy relaxation rate follows from substituting the distribution from Eq.~(\ref{eq:amax}) into Eq.~(\ref{eq:nu_def}) with Eq.~(\ref{eq:c_landau}) as the collision operator. This leads to Eq.~(\ref{eq:nrl}) with the equipartition rate~\cite{ichi:70,huba:16}
\begin{equation}
\frac{\nu_{\textrm{L}}}{\bar{\nu}} = \frac{(1+\frac{2}{3}A)^{3/2}}{A^{2}} \biggl[ - 3 + (A+3) \frac{\arctan (\sqrt{A})}{\sqrt{A}} \biggr] \ln \Lambda \label{eq:nu_nrl}
\end{equation}
where 
\begin{equation}
\bar{\nu} = 2 \sqrt{\frac{\pi}{m}} \frac{ne^4}{(k_BT)^{3/2}} = \omega_p \sqrt{\frac{3}{4\pi}} \Gamma^{3/2}
\end{equation}
is a reference collision rate. Equation~(\ref{eq:nu_nrl}) is valid for any sign of $A$, but for $A<0$ is often written in the equivalent form by replacing $\arctan (\sqrt{A})/\sqrt{A} = \arctanh ( \sqrt{-A})/\sqrt{-A}$~\cite{huba:16}. 

\begin{figure}[t]
\includegraphics[width=75mm]{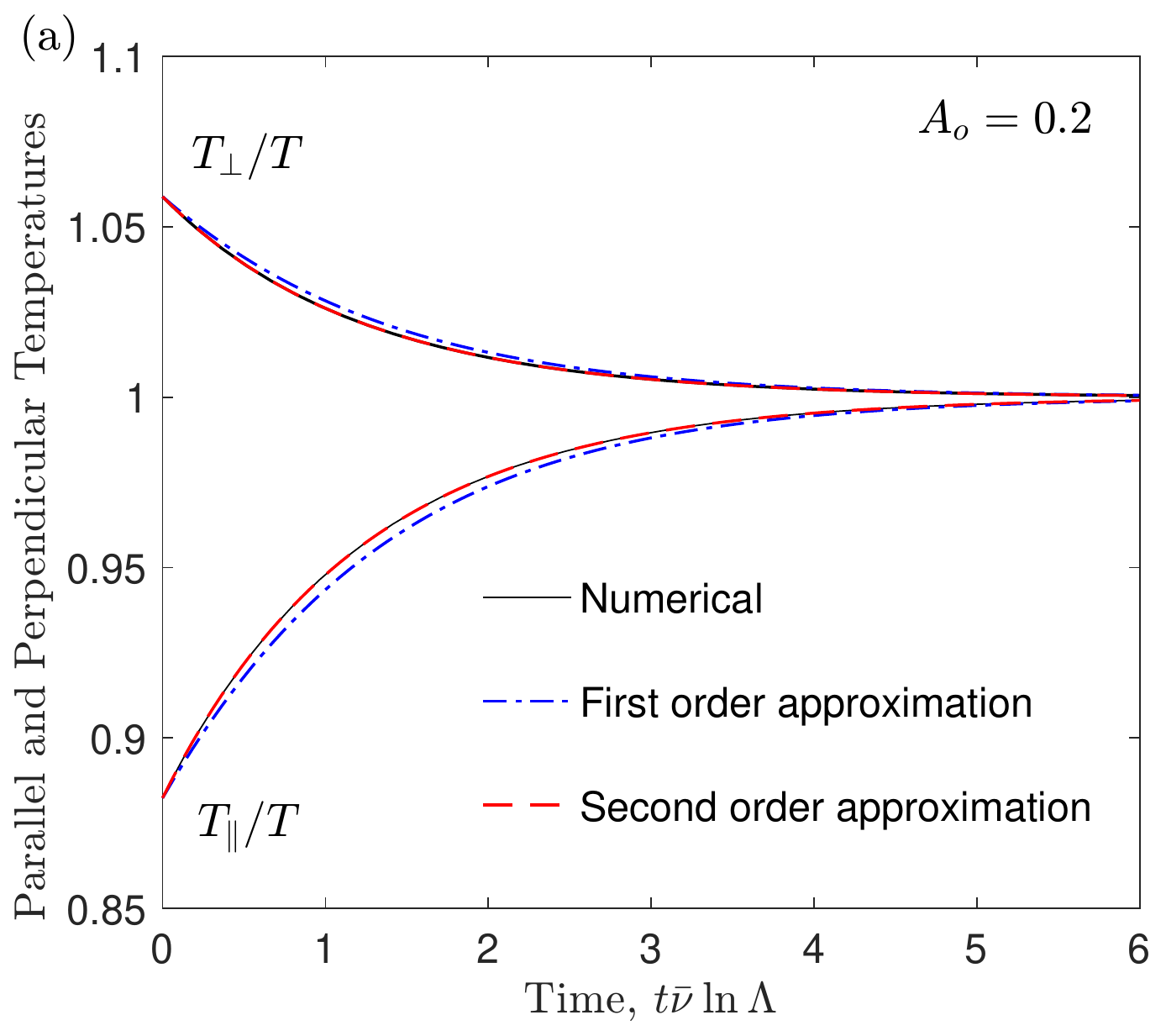}
\includegraphics[width=75mm]{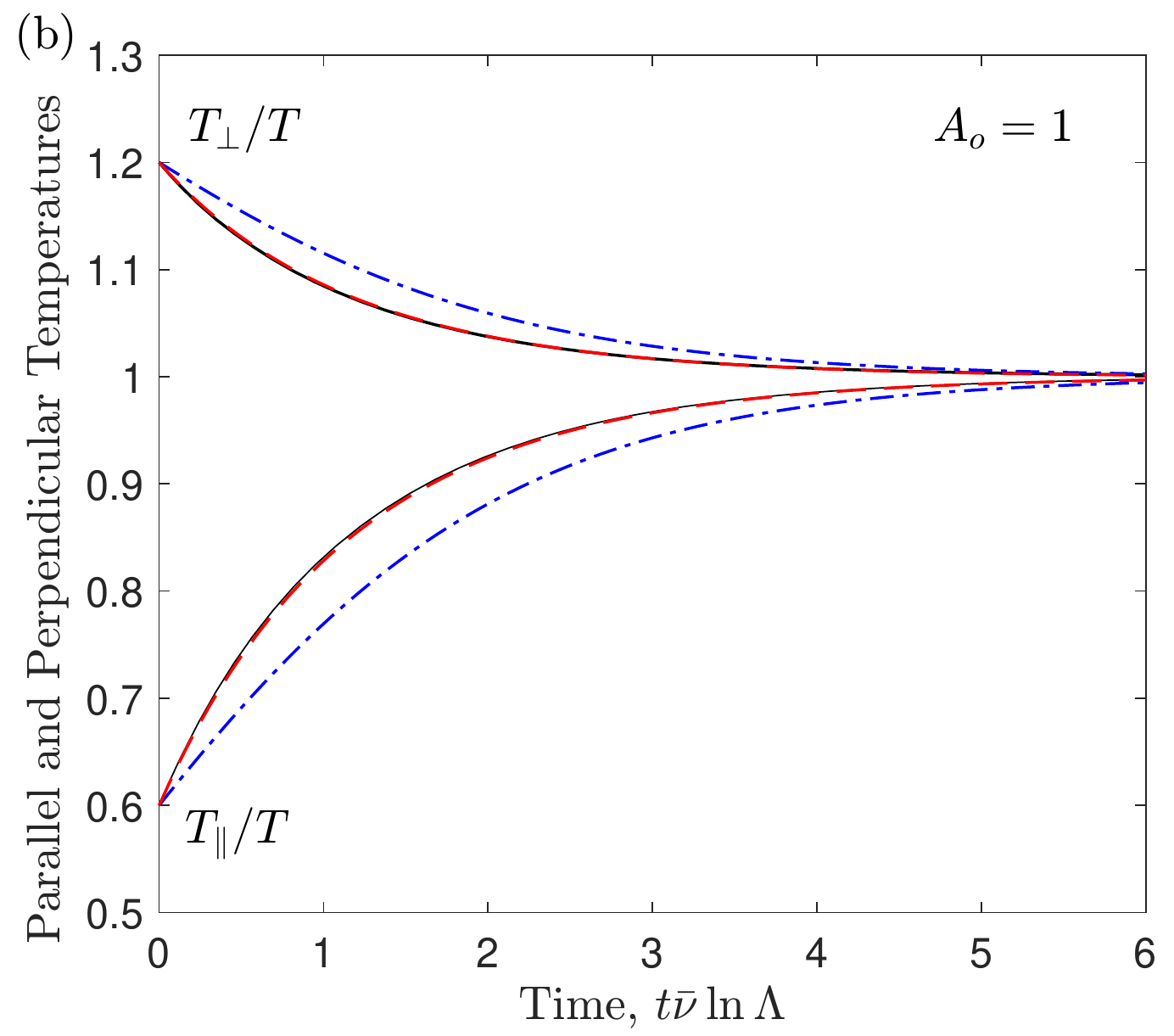}
\caption{Time evolution of $T_\parallel/T$ and $T_\perp/T$ computed from the Landau theory Eq.~(\ref{eq:nu_b0}) (solid lines), along with the first order (dash-dotted lines) and second order (dashed lines)  terms of the small anisotropy expansion from Eqs.~(\ref{eq:AL_1}) and (\ref{eq:AL_2}) respectively: (a) $A_o = 0.2$ and (b) $A_o = 1$.  }
\label{fg:T_t_landau}
\end{figure}

Equation~(\ref{eq:nu_nrl}) was derived from Eq.~(\ref{eq:c_landau}) without explicit reference to the magnitude of the anisotropy. In the weak anisotropy limit $A \ll 1$, Eqs.~(\ref{eq:dAdt}) and (\ref{eq:nu_nrl}) reduce to $dA/d\tilde{t} = -\frac{4}{5} A - \frac{68}{105} A^2 + \mathcal{O}(A^3)$, where $\tilde{t} \equiv t \bar{\nu} \ln \Lambda$. Thus, to first order in the anisotropy expansion the relaxation is exponential 
\begin{equation}
A_1(t) = A_o \exp \biggl(- \frac{4}{5} \tilde{t} \biggr) \label{eq:AL_1}
\end{equation}
and at second order has the form 
\begin{equation}
A_2(t) = \frac{21 A_o}{\exp (\frac{4}{5} \tilde{t}) (17A_o + 21)-17A_o} . \label{eq:AL_2} 
\end{equation}

Figure~(\ref{fg:T_t_landau}) shows results of the temperature evolution for initial anisotropies of $A_o = 0.2$ and $A_o=1$ predicted from Eq.~(\ref{eq:nu_nrl}). Also shown are the small anisotropy expansions from Eqs.~(\ref{eq:AL_1}) and (\ref{eq:AL_2}). This shows that for a weak initial anisotropy, the simple expression obtained from the lowest order term in the expansion, Eq.~(\ref{eq:AL_1}), provides an accurate approximation of the full result. As $A$ increases to 1, deviations from the exponential form become important, but these are accurately modeled by Eq.~(\ref{eq:AL_2}). 

\subsection{Lenard-Balescu theory}

An alternative to Landau's kinetic theory for weakly coupled plasmas is the Lenard-Balescu equation~\cite{lena:60}, which for the OCP is 
\begin{equation}
C_\textrm{LB} = - \frac{2 e^4}{m^2} \frac{\partial}{\partial \vc{v}} \cdot \int d^3v^\prime \int d^3k \frac{\delta (\vc{k} \cdot \vc{u})}{|\hat{\varepsilon} (\vc{k}, \vc{k} \cdot \vc{v})|^2} \frac{\vc{k} \vc{k}}{k^4} \cdot \biggl[ f(\vc{v}) \frac{\partial f(\vc{v}^\prime)}{\partial \vc{v}^\prime} - f(\vc{v}^\prime) \frac{\partial f(\vc{v})}{\partial \vc{v}} \biggr]  . \label{eq:c_lb}
\end{equation}
The primary distinction between this and Landau's collision operator is that the long-range screening is self-consistently accounted for via the linear dielectric response function $\hat{\varepsilon}(\vc{k}, \vc{k} \cdot \vc{v})$. In contrast to the static Debye-H\"{u}ckel screening assumed in the Landau equation, this is a generalized theory that can, in principle, account for velocity dependence (aka dynamics) of the screening. 
Following the same procedure as the previous section by using the distribution from Eq.~(\ref{eq:amax}) in Eq.~(\ref{eq:nu_def}) with Eq.~(\ref{eq:c_lb}) as the collision operator provides
\begin{equation}
\nu_{\textrm{LB}} = \frac{n e^4}{\pi} \int d^3k \int_{-\infty}^\infty d\omega \frac{1}{k^4} \frac{1}{|\hat{\varepsilon}(\vc{k}, \omega)|^2} \frac{k_\parallel^2 k_\perp^2}{(k_\parallel^2 k_BT_\parallel + k_\perp^2 k_BT_\perp)^2} \exp \biggl( - \frac{m \omega^2}{k_\parallel^2 k_BT_\parallel + k_\perp^2 k_BT_\perp} \biggr) \label{eq:nu_b0}
\end{equation}
where 
\begin{equation}
\hat{\varepsilon}(\vc{k},\omega) = 1  + \frac{\omega_p^2}{k^2} \int d^3v \frac{\vc{k} \cdot \partial f/\partial \vc{v}}{\omega - \vc{k} \cdot \vc{v}} = 1 - \frac{\omega_p^2}{k_\parallel^2 v_{T\parallel}^2 + k_\perp^2 v_{T\perp}^2} Z^\prime \biggl( \frac{\omega}{\sqrt{k_\parallel^2 v_{T\parallel}^2 + k_\perp^2 v_{T\perp}^2}} \biggr) . \label{eq:ep_hat}
\end{equation}
The expression following the second equality in Eq.~(\ref{eq:ep_hat}) follows from inserting the anisotropic Maxwellian distribution from Eq.~(\ref{eq:amax}), $Z$ is the plasma dispersion function~\cite{frie:61} and the ``prime'' denotes the first derivative with respect to the argument. 

Lenard-Balescu theory contains an ultraviolet divergence because it does not self-consistently account for the close interaction of colliding particles. This physics is imposed by limiting the minimum interaction length scale to be the classical distance of closest approach, or equivalently, the maximum wavenumber to be $k_{\textrm{max}} = 1/b_{\textrm{min}}=2k_BT/e^2$. Representing $k$ in a spherical coordinate system
 $\vc{k} = k \cos \theta \sin \phi \hat{x} + k \sin \theta \sin \phi \hat{y} + k \cos \phi \hat{z}$, Eq.~(\ref{eq:nu_b0}) can be expressed as 
\begin{equation}
\frac{\nu_{\textrm{LB}}}{\bar{\nu}} = \biggl( \frac{1+\frac{2}{3}A}{1+A} \biggr)^{3/2} \frac{1}{\sqrt{\pi}} \int_{-\infty}^\infty dx e^{-x^2} \int_{-1}^1 dy \frac{y^2 (1-y^2)}{(y^2 \tilde{A} + 1)^{3/2}} \int_0^{\bar{k}_{\max}} \frac{d \bar{k}}{\bar{k}} \biggl| 1 + \frac{\Phi(x)}{\bar{k}^2 (y^2 \tilde{A} + 1)} \biggr|^{-2} . \label{eq:nu_mid2}
\end{equation}
Here, the following variables have been defined: 
$y=\cos\phi$, $\bar{k} = k \lambda_{D\perp}$, $x = (\omega/\omega_p)/\sqrt{\bar{k}_\parallel^2 r + \bar{k}_\perp^2}$, 
$r = T_\parallel/T_\perp=1/(1+A)$, 
$\tilde{A} \equiv -A/(1+A)$, and $\Phi(x) \equiv - \frac{1}{2} Z^\prime (x/\sqrt{2})$, where $\lambda_{D\perp}^2 \equiv k_BT_\perp/(4\pi e^2 n)$ is a Debye length associated with the perpendicular temperature. 
Equation~(\ref{eq:nu_mid2}) can be further reduced by carrying out the $\bar{k}$ integral, leading to 
\begin{align}
\frac{\nu_{\textrm{LB}}}{\bar{\nu}} &= \biggl( \frac{1+\frac{2}{3}A}{1+A} \biggr)^{3/2} \frac{2}{\sqrt{\pi}} \int_{-\infty}^\infty dx e^{-x^2} \int_{0}^1 dy \frac{y^2 (1-y^2)}{(y^2 \tilde{A} + 1)^{3/2}} \biggl\lbrace  \frac{1}{4} \ln \biggl( \frac{|\bar{k}_{\max}^2 (y^2 \tilde{A}+1) + \Phi|^2}{|\Phi|^2} \biggr) \label{eq:nu_b0_full} \\ \nonumber
& - \frac{1}{2} \frac{\Phi_R}{\Phi_I} \biggl[ \arctan \biggl( \frac{\bar{k}_{\max}^2 (y^2 \tilde{A} +1) + \Phi_R}{\Phi_I} \biggr) - \arctan \biggl( \frac{\Phi_R}{\Phi_I} \biggr) \biggr] \biggr\rbrace  ,
\end{align}
where
\begin{equation}
\bar{k}_{\max} = \frac{\lambda_{D\perp}}{b_{\min}}  = \frac{\sqrt{k_BT_\perp/(4\pi e^2n)}}{e^2/(2k_BT)} = \sqrt{\frac{1+A}{1+\frac{2}{3}A}} \frac{2}{\sqrt{3} \Gamma^{3/2}},
\end{equation}
$\Phi_R = \textrm{Re} \lbrace \Phi \rbrace$ and $\Phi_I = \textrm{Im} \lbrace \Phi \rbrace$.  Equation~(\ref{eq:nu_b0_full}) follows directly from the collision operator, Eq.~(\ref{eq:c_lb}), assuming only that the distribution function remains an anisotropic Maxwellian throughout the evolution. This will be compared directly with the Landau equation result from Eq.~(\ref{eq:nu_nrl}), but first the following subsections discuss how the integrals in Eq.~(\ref{eq:nu_b0_full}) can be simplified via accurate approximations. 

\subsubsection{Isotropic Screening Approximation \label{sec:isa}}

Note that the wavenumber ($\bar{k}$) and azimuthal ($y$) integrals in Eq.~(\ref{eq:nu_mid2}) would be decoupled if not for the anisotropy of the screening (i.e., take $\tilde{A}=0$ in the last term). This feature arises in the dielectric function of Eq.~(\ref{eq:ep_hat}) because the temperature anisotropy leads to a different screening length in the parallel and perpendicular directions. However, the distance of closest approach, which is also temperature dependent, must be added in an ad-hoc way in Lenard-Balescu theory, so it is not clear what the overall anisotropy dependence of this term should be.  In any case, since the dependence of transport rates on the screening length is logarithmic in the weakly coupled limit, it has a negligible influence if the coupling parameter $\Gamma$ is sufficiently small. If the screening is assumed isotropic, one can connect the results from the Landau equation based theory, Eq.~(\ref{eq:nu_nrl}), with the results from the Lenard-Balescu equation based theory via the Coulomb logarithm. Taking $\tilde{A}=0$ in the screening term of Eq.~(\ref{eq:nu_mid2}), or equivalently in the terms in curly braces in Eq.~(\ref{eq:nu_b0_full}), we note that 
\begin{equation}
2 \int_{0}^1 dy \frac{y^2 (1-y^2)}{(y^2 \tilde{A} + 1)^{3/2}} = \frac{1}{r^{3/2}} \frac{1}{A^2} \biggl[-3 + (A+3) \frac{\arctan (\sqrt{A})}{\sqrt{A}} \biggr] .
\end{equation}
In this isotropic screening limit, the Lenard-Balescu result from Eq.~(\ref{eq:nu_b0_full}) can be written in the same form as the Landau result from Eq.~(\ref{eq:nu_nrl})
\begin{equation}
\frac{\nu_{\textrm{LB,I}}}{\bar{\nu}} = \frac{(1+\frac{2}{3}A)^{3/2}}{A^2} \biggl[ -3 + (A+3) \frac{\arctan (\sqrt{A})}{\sqrt{A}} \biggr] \Xi_{\textrm{LB,I}} \label{eq:nu_b0_approx}
\end{equation}
but where the Coulomb logarithm is replaced by 
\begin{equation}
\Xi_{\textrm{LB,I}} = \frac{1}{2} \frac{1}{\sqrt{\pi}} \int_{-\infty}^\infty dx e^{-x^2} \biggl\lbrace \ln \biggl( \frac{|\bar{k}_{\max}^2 + \Phi |}{|\Phi |} \biggr) - \frac{\Phi_R}{\Phi_I} \biggl[ \arctan \biggl( \frac{\bar{k}_{\max}^2 + \Phi_R}{\Phi_I} \biggr) - \arctan \biggl( \frac{\Phi_R}{\Phi_I} \biggr) \biggr] \biggr\rbrace . \label{eq:xi_d}
\end{equation}
This is an effective Coulomb logarithm that includes dynamic screening (via the $\Phi$ terms), as well as an $\mathcal{O}(1)$ term in addition to the logarithmic term. Section~\ref{sec:wc_comp} will show that the latter contribution is significant only when $\Gamma \gtrsim 0.1$, which also happens to be where the theory breaks down due to the onset of strong correlations. 

\subsubsection{Static Screening Limit} 

The primary difference between the Landau and Lenard-Balescu based theories is the role of dynamic screening. To quantify the role of dynamic screening within the Lenard-Balescu theory, Eq.~(\ref{eq:nu_b0_full}), we compare with the static screening limit. In this limit, $\Phi(x) = \Phi(0)$, where $\Phi_R(0) = 1$, $\Phi_I(0) = 0$ and the $x$ integral in Eq.~(\ref{eq:nu_b0_full}) can be evaluated explicitly, leading to
\begin{align}
\label{eq:nu_b0_s}
\frac{\nu_{\textrm{LB,s}}}{\bar{\nu}} &= 2 \biggl(\frac{1+\frac{2}{3}A}{1+A} \biggr)^{3/2} \int_{0}^1 dy \frac{y^2 (1-y^2)}{(y^2 (r-1) + 1)^{3/2}} \biggl[  \ln  \sqrt{\bar{k}_{\max}^2 (y^2 (r-1)+1) + 1}   \\ \nonumber
&- \frac{1}{2} \frac{\bar{k}_{\max}^2(y^2(r-1)+1)}{1 + \bar{k}_{\max}^2(y^2(r-1)+1)} \biggr]   .
\end{align}
This is compared with the full expression from Eq.~(\ref{eq:nu_b0_full}) in Fig.~\ref{fg:nu_wc_comp} (discussion in Sec.~\ref{sec:wc_comp}).

\subsubsection{Debye-H\"{u}ckel Screening\label{sec:dh}} 

Finally, to connect the Lenard-Balescu and Landau results, the Deybe-H\"{u}ckel screening limit is considered. The linear dielectric corresponding to Debye-H\"{u}ckel screening is $\hat{\varepsilon} = 1 + 1/(k \lambda_D)^2 = 1 + 3/(\bar{k}^2(2+r))$. Using this in Eq.~(\ref{eq:nu_b0}) leads to the same form for the relaxation rate as Eqs.~(\ref{eq:nu_nrl}) and (\ref{eq:nu_b0_approx}), 
\begin{equation}
\label{eq:nu_b0_dh}
\frac{\nu_{\textrm{DH}}}{\bar{\nu}} = \frac{(1+\frac{2}{3}A)^{3/2}}{A^2} \biggl[ -3 + (A+3) \frac{\arctan (\sqrt{A})}{\sqrt{A}} \biggr] \Xi_\textrm{DH}
\end{equation}
but where
\begin{equation}
\Xi_\textrm{DH} = \ln \bigl( \sqrt{\Lambda^2 + 1} \bigr) - \frac{1}{2} \frac{\Lambda^2}{\Lambda^2 + 1} . \label{eq:xi_s}
\end{equation}
Note that for $\Lambda \gg 1$, $\Xi_{\textrm{DH}} \rightarrow \ln \Lambda + \mathcal{O}(1)$, returning the conventional result from Eq.~(\ref{eq:nu_nrl}). 

\subsection{Comparison of weakly coupled plasma theories \label{sec:wc_comp}} 

\begin{figure}[t]
\includegraphics[width=75mm]{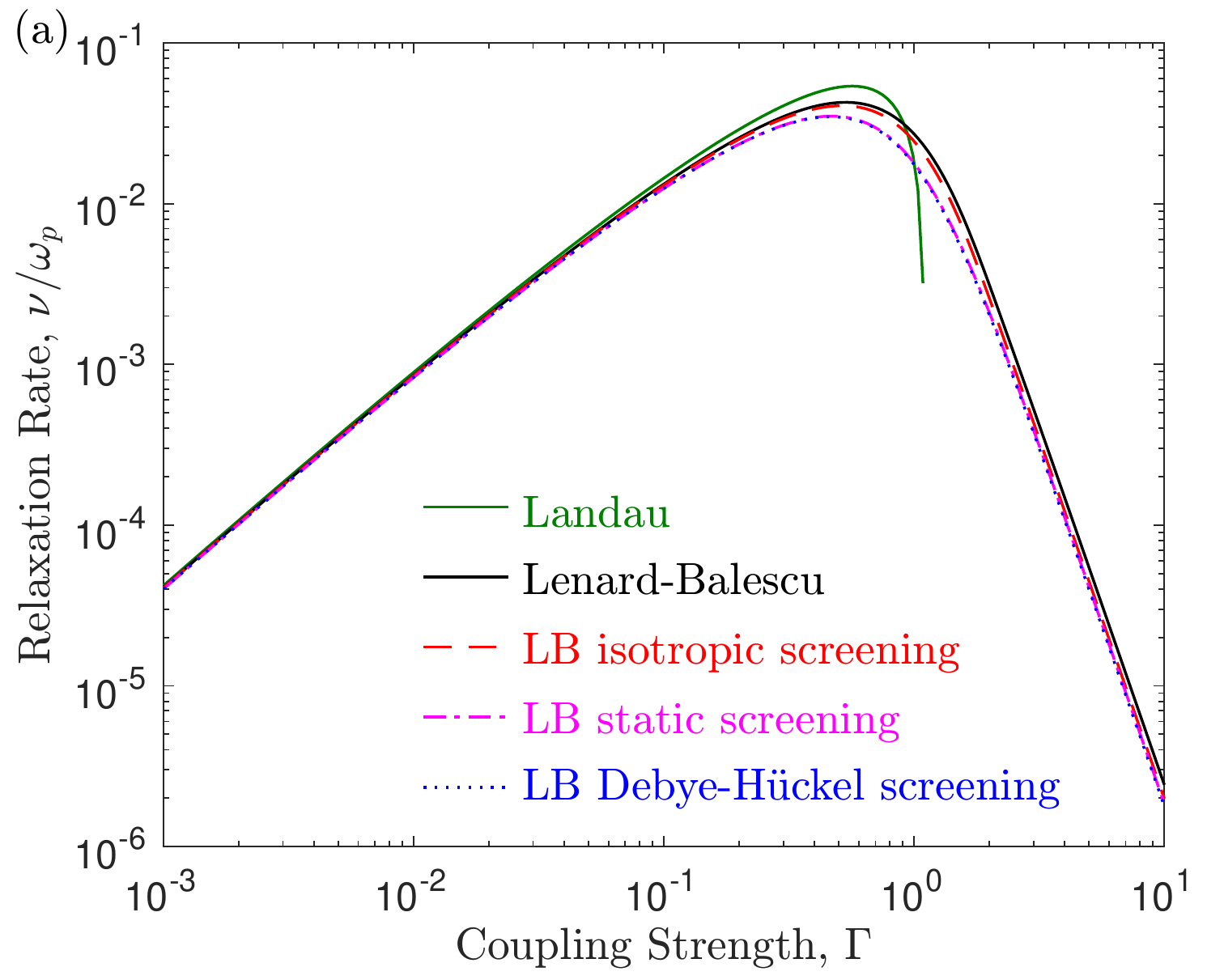}
\includegraphics[width=73mm]{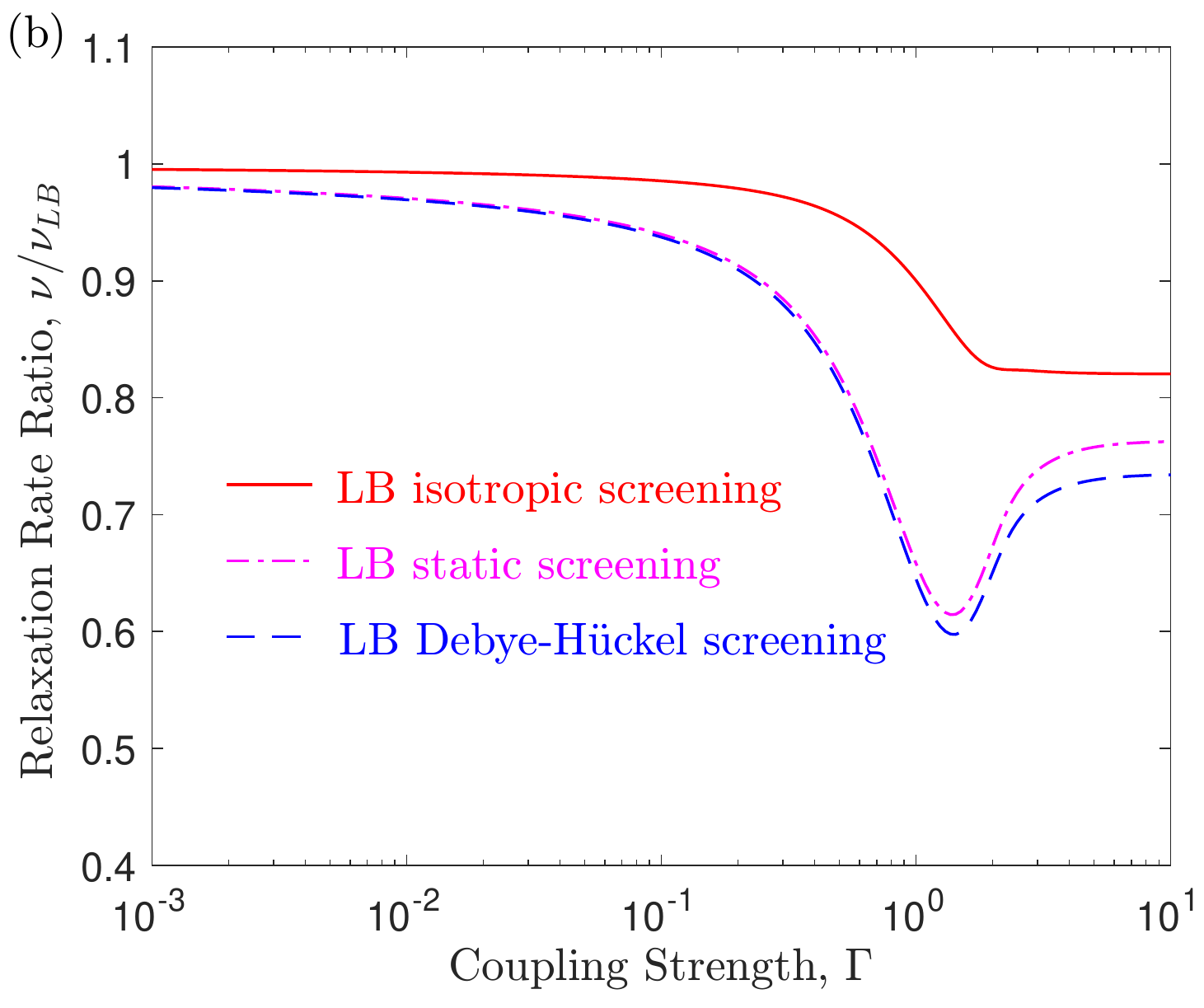}
\caption{All figures use $T_\perp/T_\parallel = 0.8$ ($A=-0.2$): (a) Relaxation rate computed from the Landau theory from Eq.~(\ref{eq:nu_nrl}) (green solid line) and the Lenard-Balescu theory from Eq.~(\ref{eq:nu_b0_full}) (black solid line). Also shown are various limits of the Lenard-Balescu result: the isotropic screening result from Eq.~(\ref{eq:nu_b0_approx}) (red dashed line), the static screening approximation from Eq.~(\ref{eq:nu_b0_s}) (magenta dash-dotted line), and the Debye-H\"{u}ckel screening model from Eq.~(\ref{eq:nu_b0_dh}) (blue dotted line). (b) Ratio of approximate solutions of the Lenard-Balescu result from Eqs.~(\ref{eq:nu_b0_approx}), (\ref{eq:nu_b0_s}) and (\ref{eq:nu_b0_dh}) to the full numerical solution from Eq.~(\ref{eq:nu_b0_full}).}
\label{fg:nu_wc_comp}
\end{figure}

Figure~\ref{fg:nu_wc_comp} shows a comparison of the results of the weakly coupled theories described in this section. Panel (a) shows that both the Landau result from Eq.~(\ref{eq:nu_nrl}) and the Lenard-Balescu result from Eq.~(\ref{eq:nu_b0_full}) give essentially indistinguishable predictions for $\Gamma \lesssim 0.1$ at $A=-0.2$ (a weak temperature anisotropy $T_\perp/T_\parallel = 0.8$). This is because dynamic screening leads to a $\mathcal{O}(1)$ correction to the Coulomb logarithm; only super-thermal particles exhibit significant dynamic screening, but this transport process is controlled by thermal particles. Since $\ln \Lambda$ becomes large in the weakly coupled regime, any contribution from dynamic screening is inconsequential. For $\Gamma \gtrsim 0.1$ the two results differ, and the predictions of the Landau theory becomes negative near $\Gamma \simeq 1$. However, we will see in the next section that neither of these theories are accurate for $\Gamma \gtrsim 0.1$ because correlation effects onset, which require a strongly coupled plasma theory to describe. Thus, there seems to be no advantage to using the comparatively complicated Lenard-Balescu theory to describe this process. Figure~\ref{fg:nu_wc_comp}b shows the ratio of each of the approximations of the Lenard-Balescu result with the full numerical result. The results of the static screening model, and the Debye-H\"{u}ckel screening model are within a few percent of the full results for $\Gamma \lesssim 0.1$; further emphasizing the conclusion that dynamic screening does not significantly influence this process. The isotropic screening assumption is excellent for $\Gamma \lesssim 1$, beyond which the Lenard-Balescu theory is not expected to be accurate. The strongly coupled regime requires a theory that can account for correlation effects, which is the topic of the next section. 

\begin{figure}[t]
\sidecaption
\includegraphics[width=80mm]{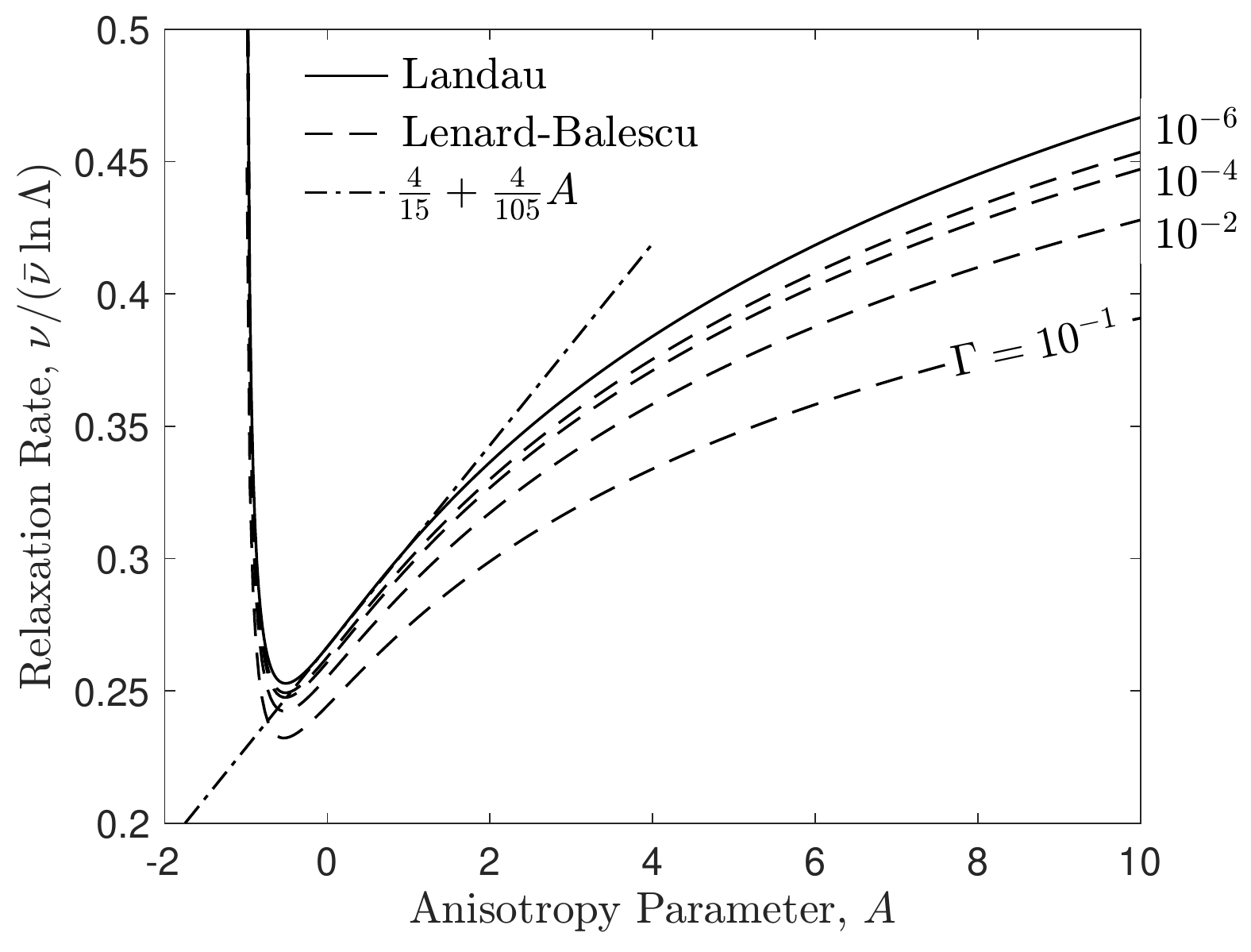}
\caption{Relaxation rate from the Landau theory of Eq.~(\ref{eq:nu_nrl}) (solid line) and the Lenard-Balescu theory of Eq.~(\ref{eq:nu_b0_full}) (dashed lines) as a function of the anisotropy parameter $A$. The linear expansion of Eq.~(\ref{eq:nu_nrl}), $\nu/(\bar{\nu}\ln \Lambda) = 4/15 + 4/105 A$ (dash-dotted line) is also shown.  The relaxation rate is given in units of $\bar{\nu}\ln \Lambda$, so the Landau theory takes a single curve, whereas results of the Lenard-Balescu theory are shown at four values of $\Gamma = 10^{-1}$, $10^{-2}$, $10^{-4}$ and $10^{-6}$. }
\label{fg:nu_A_comp}
\end{figure}

Figure~\ref{fg:nu_wc_comp} shows that the influence of dynamic screening is negligible when the anisotropy is small. However, it may be expected to become more important as the anisotropy increases. Figure~\ref{fg:nu_A_comp} shows how the relaxation rate predicted by each theory depends on the anisotropy parameter $A$. The relaxation rate is plotted in units of $\bar{\nu} \ln \Lambda$, so the Landau result from Eq.~(\ref{eq:nu_nrl}) provides a single curve. The Taylor expansion of this curve up to linear order, $\nu/(\bar{\nu}\ln \Lambda) = 4/15 +(4/105)A$, shows that, as expected, the expansion holds in a narrow neighborhood near $A=0$. The accuracy of the expansion extends much further on the $A>0$ side than for $A<0$. The dashed lines show the predictions of the Lenard-Balescu result from Eq.~(\ref{eq:nu_b0_full}) at $\Gamma = 10^{-1}$, $10^{-2}$, $10^{-4}$ and $10^{-6}$. Comparison of these curves shows that, indeed, dynamic screening has a larger influence as the anisotropy increases or the coupling strength increases. For example, at $A=10$ and $\Gamma = 10^{-2}$ there is an approximately 10\% difference between the Landau and Lenard-Balescu predictions. Two notes should be kept in mind when interpreting this result. 

First, the Lenard-Balescu theory treats the long-distance behavior of Coulomb interactions, which corrects the infrared divergence of Landau theory and incorporates temperature anisotropy in the screening length. However, it still contains an ultraviolet divergence, which is resolved by limiting the distance of closest approach to be the thermal Landau length $r_L = e^2/(2k_BT)$. The temperature arising in this length scale is assumed to be the total temperature, but it is not clear if it should also account for temperature anisotropy. 

Second, the potential role of dynamic screening should be analyzed keeping in mind that both theories are making an assumption that the distribution function remains an anisotropic Maxwellian throughout the evolution. Expectations from other transport processes, such as diffusion, suggest that the error in this approximation is on the order of 20\% at weak coupling~\cite{ferz:72,baal:14}. Some evidence for this from comparison with MD simulations will also be shown in Sec.~\ref{sec:results}. Thus, the error associated with approximations related to the distribution function are often more significant than changes in the transport rates associated with anisotropy in the screening.

\section{Strongly Coupled Plasma Theories\label{sec:sc}}

Both the Landau and Lenard-Balescu based theories from the previous section address the weakly coupled limit. In this section, two theories are evaluated that attempt to extend these methods into the strongly coupled regime. Each of these is based on phenomenological assumptions, rather than rigorous asymptotic expansions, so the subsequent comparison with MD serves as a test of these assumptions. 

\subsection{Effective Potential Theory}

Like the Landau approach, the effective potential theory (EPT)~\cite{baal:13,baal:14,baal:15} is based on the Boltzmann collision operator, which for the OCP is
\begin{equation}
C_\textrm{B} = \int d^3v^\prime \int d\Omega\, \sigma\, u [f(\hat{\vc{v}}) f(\hat{\vc{v}}^\prime) - f(\vc{v}) f(\vc{v}^\prime)], \label{eq:c_boltz}
\end{equation}
where $\sigma$ is the differential scattering cross section, the hat denotes the velocity vector after a binary collision: $\hat{\vc{v}} = \vc{v} + \Delta \vc{v}$ and $\vc{u} \equiv \vc{v} - \vc{v}^\prime$. Unlike the other theories discussed in this paper, the anisotropy relaxation rate has not previously been calculated from this theory, so we provide details of the calculation here. Starting with the $v_\parallel^2$ moment of Eq.~(\ref{eq:c_boltz}) provides
\begin{equation}
\frac{dT_\parallel}{dt} = \frac{m}{n} \int d^3v\ v_\parallel^2 C_B(f) = \frac{m}{n}  \int d^3v \int d^3v^\prime \lbrace \Delta (v_\parallel^2) \rbrace f(\vc{v}) f (\vc{v}^\prime) \label{eq:ept_1}
\end{equation}
where the curly brackets are defined as $\lbrace \chi \rbrace = \int d\Omega \sigma u \Delta \chi$~\cite{baal:12}, for any function $\chi$, $\Delta (v_\parallel^2) = \hat{v}_\parallel^2 - v_\parallel^2$, and momentum conservation implies $\Delta \vc{v} = \Delta \vc{u}/2$, so $\Delta (v_\parallel^2) = v_\parallel \Delta u_\parallel + \frac{1}{4} (\Delta u_\parallel)^2$. 

Assuming that the distribution function has the anisotropic Maxwellian form described by Eq.~(\ref{eq:amax}), the three velocity integrals for the variable $\vc{v}$ in Eq.~(\ref{eq:ept_1}) can be computed analytically leading to
\begin{equation}
\frac{dT_\parallel}{dt} = \frac{n}{\pi^{3/2} v_{T\perp}^2 v_{T\parallel}} \frac{m}{2^{9/2}} \int d^3u\, u (u^2 - 3 u_\parallel^2) \bar{\sigma}^{(2)}(u) \exp \biggl(- \frac{1}{2} \frac{u_\perp^2}{v_{T\perp}^2} \biggr) \exp \biggl( - \frac{1}{2} \frac{u_\parallel^2}{v_{T\parallel}^2} \biggr) \label{eq:ept_2}
\end{equation}
in which 
\begin{equation}
\bar{\sigma}^{(l)} = 2\pi \int_0^\infty db\, b[1 - \cos^l (\pi - 2 \Theta)] \label{eq:sig_l}
\end{equation}
is the $l^\textrm{th}$ momentum scattering cross section and $\theta = \pi - 2\Theta$ is the scattering angle, where
\begin{equation}
\Theta = b \int_{r_o}^\infty dr\, r^{-2} [1-b^2/r^2 - (e\phi(r)/k_BT)/\xi^2]^{-1/2}  . \label{eq:theta}
\end{equation}
Obtaining Eq.~(\ref{eq:ept_2}) from Eq.~(\ref{eq:ept_1}) makes use of the relations $\lbrace \Delta \vc{u} \rbrace = -u \vc{u} \bar{\sigma}^{(1)} (u)$ and $\lbrace \Delta \vc{u} \Delta \vc{u} \rbrace = u[2 \vc{u} \vc{u} \bar{\sigma}^{(1)} + \frac{1}{2} (u^2 \ident - 3 \vc{u} \vc{u}) \bar{\sigma}^{(2)} (u) ]$, which imply $u_\parallel \lbrace \Delta u_\parallel \rbrace + \frac{1}{2} \lbrace \Delta (u_\parallel^2) \rbrace = \frac{1}{4} u (u^2 - 3u_\parallel^2) \bar{\sigma}^{(2)} $. Applying a spherical coordinate system $\vc{u} = u[\cos \theta^\prime\, \sin\phi^\prime \hat{x} + \sin \theta^\prime \sin \phi^\prime \hat{y} + \cos \phi^\prime \hat{z}]$ both the $\theta^\prime$ and $\phi^\prime$ integrals can be carried out. This leads to Eq.~(\ref{eq:nrl}) where the anisotropy relaxation rate is 
\begin{equation}
\frac{\nu_\textrm{EPT}}{\bar{\nu}} = \chi \frac{3 \sqrt{\pi}}{16} \frac{(1+\frac{2}{3}A)^{3/2}}{\sqrt{\alpha} A^{5/2}} \int_0^\infty d\xi\, \xi^2 e^{-\alpha \xi^2} \frac{\bar{\sigma}^{(2)}}{\sigma_o} \biggl[ \frac{2}{3} \xi^2 \alpha A \erf (\xi \sqrt{\alpha A}) - \psi (\xi^2 \alpha A) \biggr] .  \label{eq:nu_ept}
\end{equation}
Here, $\sigma_o \equiv \pi e^4/(2k_BT)^{2}$, $\xi^2 \equiv \frac{1}{2} u^2/v_{T}^2 = u^2/(4k_BT/m)$, $\alpha \equiv T/T_\perp =  \frac{1}{3} (3+A)/(1+A)$
 and
\begin{equation}
\psi(x) \equiv \erf( \sqrt{x}) - \frac{2}{\sqrt{\pi}} \sqrt{x} e^{-x}
\end{equation}
is the Maxwell integral. The right side of Eq.~(\ref{eq:nu_ept}) has been multiplied by a factor $\chi$, which is an aspect of EPT that will be explained below. 

The expression in Eq.~(\ref{eq:nu_ept}) follows directly from inserting the anisotropic distribution function in the Boltzmann collision operator. The EPT extends this Boltzmann-based result by relaxing (but not eliminating) two of its underlying assumptions: binary collisions and molecular chaos. First, the main premise of the theory is that binary collisions between particles do not occur in isolation, but rather in a ``sea'' of background particles. This ``relaxes'' Boltzmann's assumption by accounting for aspects of many-body physics via the interaction potential. The appropriate potential is one which takes into account the fields associated with nearby charges. In particular, the potential of mean force is the potential obtained when taking two particles at fixed positions and averaging over the positions of all other particles~\cite{hill:60}
\begin{eqnarray}
\vc{F}_{12} &=& \int \biggl[ - \nabla_{\vc{r}_1} U(\vc{r}_1, \vc{r}_2, \ldots , \vc{r}_N) \biggr] \frac{e^{-U/k_BT}}{\mathcal{Z}} d\vc{r}_3 \ldots d\vc{r}_N  \label{eq:f12} \\ \nonumber
&=& k_BT \nabla_{\vc{r}_1} \ln g(|\vc{r}_1 - \vc{r}_2|) \equiv - \nabla_{\vc{r}_1} \phi (\vc{r}_1 - \vc{r}_2) .
\end{eqnarray}
Here $\mathcal{Z} = \int \exp(-U/k_BT) d\vc{r}_N$ is the configurational integral and $U \equiv \sum_{i,j} v(|\vc{r}_i - \vc{r}_j|)$.  This depends only on the bare interaction potential of the particles $v_{ij}$, which is the Coulomb potential. Thus, once $g(r)$ is determined, the interaction potential is modeled as $\phi(r) = - k_BT \ln[g(r)]$. 

Second, the factor $\chi$ in Eq.~(\ref{eq:nu_ept}) accounts for an increased collision rate due to the excluded volume surrounding particles interacting via a repulsive (Coulomb) potential~\cite{baal:15}. This excluded volume is sometimes referred to as the Coulomb hole. It decreases the volume of space that particles can occupy, resulting in an increased scattering rate. Accounting for this relaxes Boltzmann's molecular chaos assumption somewhat because it introduces an aspect of correlation in the initial positions of scattering particles. We model this factor using a modified version of Enskog's theory of hard sphere gases that has been adapted to plasmas~\cite{baal:15}. In this theory, the collision frequency enhancement factor is 
\begin{equation}
\chi \simeq 1 + 0.6250 b \rho + 0.2869 (b\rho)^2 \label{eq:chi}
\end{equation}
where $b\rho \simeq \pi n \bar{\sigma}^3/3$ and $\bar{\sigma}$ is the effective particle diameter. This is determined from the location where $g(r=\bar{\sigma}) = 0.87$. For the OCP at $\Gamma \lesssim 1$, $\chi \simeq 1$, while for $\Gamma \simeq 1-100$, $\chi \simeq 1.2-1.4$. This 20-40\% increase in the collision frequency  can be seen in figure 7 of \cite{baal:15}. 
 
Equations~(\ref{eq:sig_l})--(\ref{eq:nu_ept}) along with $\phi = -k_BT \ln [g(r)]$ and (\ref{eq:chi}) provide a closed set of equations based on one input: $g(r)$. To compute $g(r)$ we use the hypernetted chain (HNC) approximation~\cite{hans:06}
\begin{subequations}
\label{eq:hnc}
\begin{align}
g(r) &= \exp[- v(r)/k_BT + h(\vc{r}) - c(\vc{r}) + b(\vc{r})] \\ 
\hat{h} (\vc{k}) &= \hat{c}(\vc{k}) [1 + n\hat{h}(\vc{k})]
\end{align}
\end{subequations}
where $v(\vc{r})/k_BT = \Gamma a/r$ is the bare Coulomb potential, $h(\vc{r}) \equiv g(\vc{r}) - 1$ and ``hats'' denote Fourier transforms in the spatial coordinate. In Eq.~(\ref{eq:hnc}), $b(\vc{r})$ is a bridge function that that increases the accuracy of the HNC approximation at high $\Gamma$. For this term, we apply the model of Iyetomi \textit{et al}\cite{iyet:92}. The HNC approximation has long been benchmarked for the Coulomb OCP, and it has been shown to provide accurate input for the EPT over the range of coupling strengths considered here~\cite{baal:13,baal:14}.  

Finally, we note that by modeling the interaction potential using an equilibrium theory, we have implicitly assumed that the effective interaction potential is spherically symmetric. For a large enough temperature anisotropy this approximation should be expected to break down. Next, we consider the weak anisotropy limit and the weakly coupled limit of the EPT theory.

\subsubsection{Weak anisotropy expansion} 

If the temperature anisotropy is small, $|A| \ll 1$, expanding Eq.~(\ref{eq:nu_ept}) gives
\begin{equation}
\frac{\nu}{\bar{\nu}} = \frac{2}{15} \Xi^{(2,2)} + A \biggl( \frac{2}{63} \Xi^{(2,3)} - \frac{2}{45} \Xi^{(2,2)} \biggr) + \mathcal{O}(A^{3/2})  \label{eq:nu_ept_sa}
\end{equation}
in which the generalized Coulomb logarithms 
\begin{equation}
\Xi^{(l,k)} = \frac{\chi}{2} \int_0^\infty d\xi\, \xi^{3k+3} e^{-\xi^2} \bar{\sigma}^{(l)}/\sigma_o
\end{equation}
have been identified in the analogous way as Refs.~\cite{baal:12,baal:13}. Equation~(\ref{eq:nu_ept_sa}) returns the same result as the small anisotropy expansions of the Landau theory from Eq.~(\ref{eq:nu_nrl}) if the appropriate Coulomb logarithm is substituted for $\ln \Lambda$. 
This can be obtained by applying the weakly coupled limit of the generalized Coulomb logarithms from \cite{baal:14}, $\Xi^{(l,k)} = l \Gamma(k) \ln \Lambda$, where $\Gamma(k)$ is the Gamma function.

\subsubsection{Weakly coupled limit} 

To connect Eq.~(\ref{eq:nu_ept}) with the standard result from weakly coupled plasma theory, Eq.~(\ref{eq:nu_nrl}), consider taking the weak coupling limit of Eq.~(\ref{eq:nu_ept}).  In this limit, $\bar{\sigma}^{(l)} = \lambda_D^2 4\pi l \ln (\Lambda \xi^2)/(\Lambda \xi^2)^2$ \cite{baal:14}, and the $\xi$ dependence of the Coulomb logarithm is a small correction because $\Lambda$ is large, so $\bar{\sigma}^{(l)}/\sigma_o \simeq 4 l \ln \Lambda/\xi^4$. With this, Eq.~(\ref{eq:nu_ept}) is 
\begin{equation}
\frac{\nu}{\bar{\nu}} = \frac{3 \sqrt{\pi}}{2} \ln \Lambda \frac{(1+\frac{2}{3}A)^{3/2}}{A^{5/2}} \biggl[ \frac{2}{3} A \int_0^\infty dx\, e^{-x^2} \erf (x \sqrt{A}) - \int_0^\infty dx\, e^{-x^2} \frac{\psi(x^2A)}{x^2} \biggr] .
\end{equation}
The first of these integrals can be written in terms of the $\arctan$ function $\int_0^\infty dx\, e^{-x^2} \erf(x \sqrt{A}) =\arctan (\sqrt{A})/\sqrt{\pi}$. The second integral can be evaluated by first integrating by parts, then identifying a similar integral that can be evaluated in terms of $\arctan$. The result returns Eq.~(\ref{eq:nu_nrl}). 
In the weak anisotropy limit $|A| \ll 1$,  $\nu/\bar{\nu} \rightarrow [4/15 + (4/105) A + \mathcal{O}(A^{3/2})] \ln \Lambda$. Alternatively, this can be obtained from Eq.~(\ref{eq:nu_ept_sa}) by applying the weakly coupled limit of the generalized Coulomb logarithms.

\subsection{Generalized Lenard-Balescu Theory\label{sec:glb}}

The second theory that we consider is Ichimaru's generalization of the Lenard-Balescu equation~\cite{ichi:92}
\begin{equation}
C_\textrm{I} = - \frac{2 e^4}{m^2} \frac{\partial}{\partial \vc{v}} \cdot \int d^3v^\prime \int d^3k \frac{[1-G(k)] \delta (\vc{k} \cdot \vc{u})}{|\hat{\varepsilon}_\textrm{I} (\vc{k}, \vc{k} \cdot \vc{v})|^2} \frac{\vc{k} \vc{k}}{k^4} \biggl[ f (\vc{v}) \frac{\partial f(\vc{v}^\prime)}{\partial \vc{v}^\prime} - f(\vc{v}^\prime) \frac{\partial f(\vc{v})}{\partial \vc{v}} \biggr]  . \label{eq:c_ichi}
\end{equation}
Here, $G(k)$ is the local field correction (LFC) which accounts for static correlations. For the OCP~\cite{ichi:92}  
\begin{equation}
1 - G(k) = \frac{(ka)^2}{3 \Gamma} \biggl( \frac{1}{S(k)} - 1 \biggr)
\end{equation}
where 
\begin{equation}
S(k) = 1 + \frac{3}{ka^3} \int_0^\infty dr\, r \sin (kr) [g(r) - 1]
\end{equation}
is the static structure factor. This will be determined by the HNC approximation described in Eq.~(\ref{eq:hnc}). Inserting the anisotropic distribution function from Eq.~(\ref{eq:amax}) leads to the evolution equation Eq.~(\ref{eq:nu_def}) with the relaxation rate given by
\begin{equation}
\nu_{\textrm{I}} = \frac{n e^4}{\pi} \int d^3k \int_{-\infty}^\infty d\omega \frac{1-G(k)}{k^4 |\hat{\varepsilon}_\textrm{I}(\vc{k}, \omega)|^2} \frac{k_\parallel^2 k_\perp^2}{(k_\parallel^2 k_BT_\parallel + k_\perp^2 k_BT_\perp)^2} \exp \biggl( - \frac{m \omega^2}{k_\parallel^2 k_BT_\parallel + k_\perp^2 k_BT_\perp} \biggr) \label{eq:nu_ichi}
\end{equation}
where 
\begin{equation}
\hat{\varepsilon}_\textrm{I}(\vc{k},\omega) = 1  + \frac{\omega_p^2 [1-G(k)]}{k^2} \int d^3v \frac{\vc{k} \cdot \partial f/\partial \vc{v}}{\omega - \vc{k} \cdot \vc{v}} = 1 - \frac{\omega_p^2 [1-G(k)]}{k_\parallel^2 v_{T\parallel}^2 + k_\perp^2 v_{T\perp}^2} Z^\prime \biggl( \frac{\omega}{\sqrt{k_\parallel^2 v_{T\parallel}^2 + k_\perp^2 v_{T\perp}^2}} \biggr)  \label{eq:ep_ichi}
\end{equation}
is the dielectric function. 

Unlike the Lenard-Balescu equation, Eq.~(\ref{eq:c_ichi}) does not suffer an ultraviolet divergence because the local field correction accounts for the close interaction physics. For computational simplicity, one of the $k$ integrals can be computed in an analogous way to what was done in Eq.~(\ref{eq:nu_mid2}) providing a similar result that includes the LFC 
\begin{equation}
\frac{\nu_{\textrm{I}}}{\bar{\nu}} = \biggl( \frac{1+\frac{2}{3}A}{1+A} \biggr)^{3/2} \frac{1}{\sqrt{\pi}} \int_{-\infty}^\infty dx e^{-x^2} \int_{-1}^1 dy \frac{y^2 (1-y^2)}{(y^2 \tilde{A} + 1)^{3/2}} \int_0^{\infty} d \bar{k} \frac{ [1-G(\bar{k})}{\bar{k}} \biggl| 1 +  \frac{\Phi(x)[1-G(\bar{k})]}{\bar{k}^2(y^2 \tilde{A} + 1)} \biggr|^{-2} . \label{eq:glb_mid2}
\end{equation}
Limiting values of Eq.~(\ref{eq:glb_mid2}) that are similar to what is listed for the Lenard-Balescu equation in Secs.~\ref{sec:isa}--\ref{sec:dh} can also be taken, providing similar results but where the LFC enters the expressions for the generalized Coulomb logarithm in place of the cutoff $b_{\min}$. The full expression is compared with the EPT predictions in Sec.~\ref{sec:results}.

\section{MD Simulation Description\label{sec:md}} 


\begin{figure}[t]
\includegraphics[width=75mm]{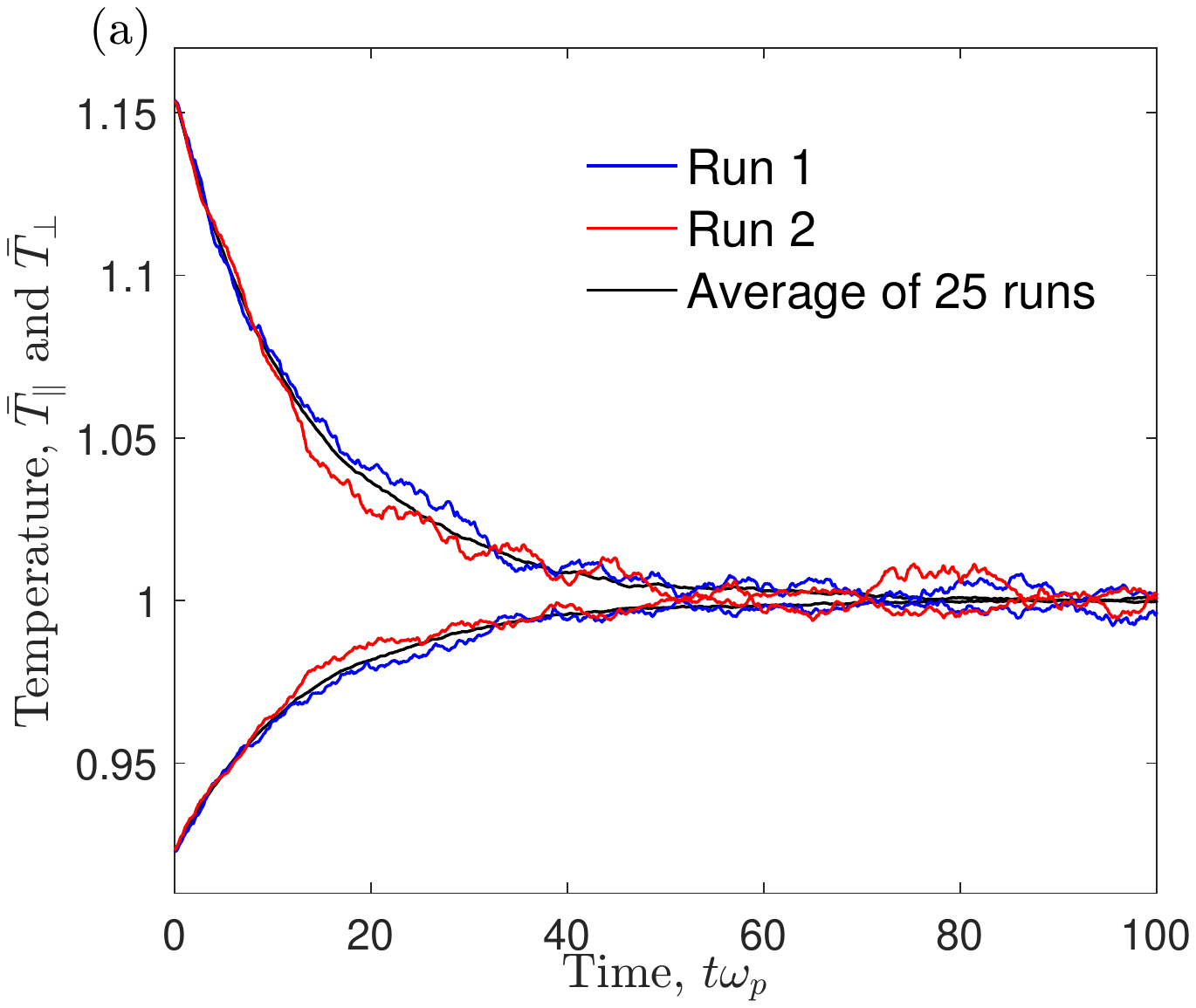}
\includegraphics[width=75mm]{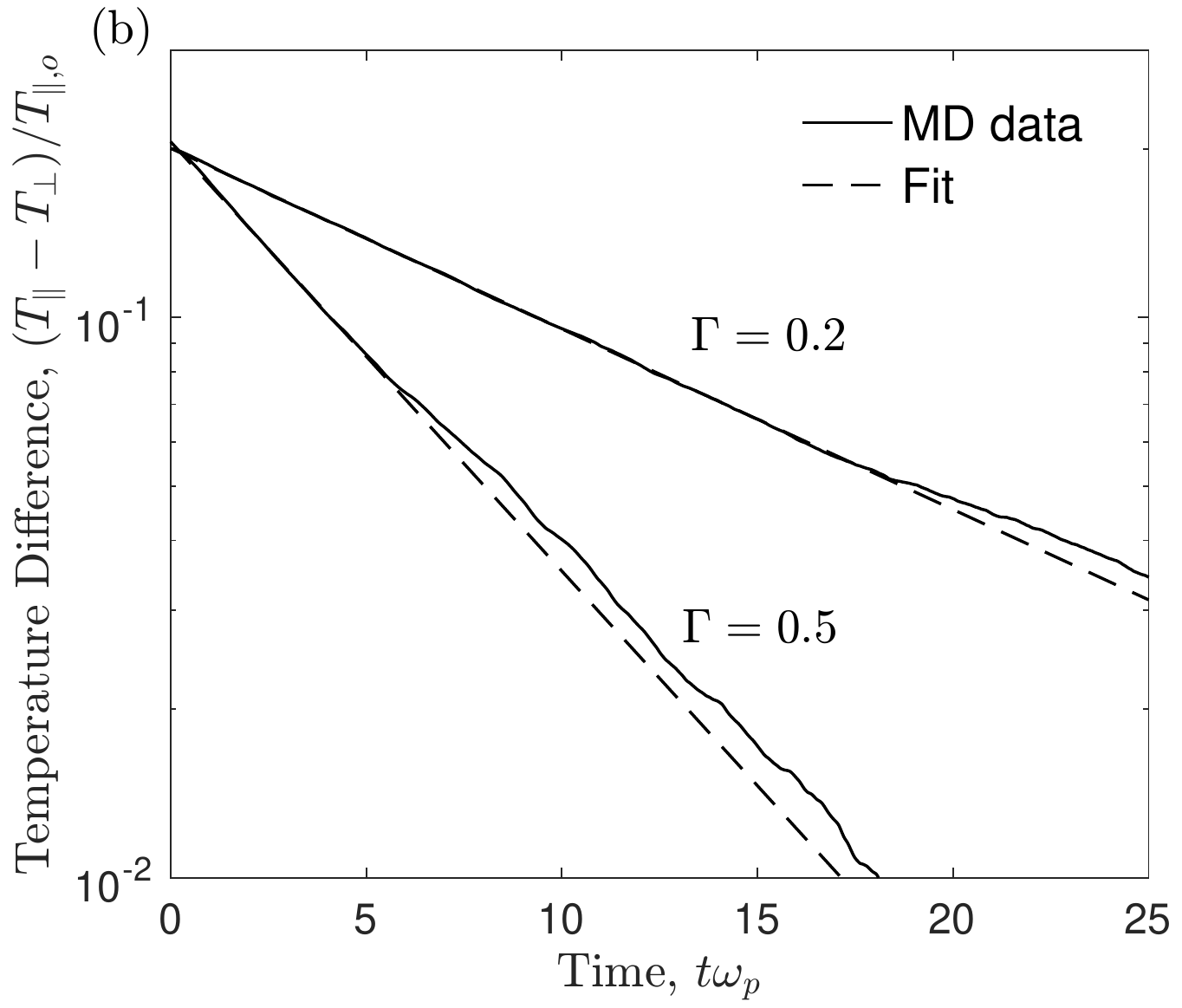}
\caption{(a) Parallel and perpendicular temperature profiles in time computed from two independent MD runs (blue and red lines), as well as the result obtained from the average of $25$ independent runs (black line). In these simulations $\Gamma = 0.2$ and $A=-0.2$. (b) Difference in the anisotropic temperatures in time from MD simulations (solid lines) and exponential fits over a time interval (dashed lines). Here, $T_{\parallel,o} = 1.15$.}
\label{fg:md_analysis}
\end{figure}
The MD simulations were performed as follows.
$N$ charged particles interacting through the pure Coulomb interaction in a uniform, neutralizing background were placed in a cubic box of volume $V$.
Periodic conditions were imposed on all boundaries.
Particle trajectories were determined by solving Newton's equations of motion with the velocity Verlet integrator \cite{FrenkelSmit}.
The force on an ion that results from its interaction with the ions in the simulation box and with those in the periodically replicated cells were calculated using the Ewald summation technique.
For numerical efficiency, the Ewald sum was calculated with a parallel implementation of the particle-particle-particle-mesh ($\rm P^3M$) method that simultaneously provides high resolution for individual encounters combined with rapid, mesh-based, long range force calculations \cite{HockneyEastwood}.
Time was normalized to the plasma frequency $\omega_p$.
The integration time step $\delta t$ and the number of particles $N$ given below were chosen in order to conserve energy to $<10^{-5}$, and to ensure high enough collision ability in the simulation cell (the calculations are more demanding at small coupling due to long collision mean-free path).
Specifically, for $\Gamma\leq 0.1$, we used $N=10^5$ and $\delta t=10^{-3}/\omega_p$; for $0.1<\Gamma<1$, $N=50000$ and $\delta t=0.01/\omega_p$; for $\Gamma\geq 1$, $N=5000$ and $\delta=0.01/\omega_p$.
 
Each simulation consisted of an equilibration phase of length $t_{\textrm{eq}}=N_{\textrm{eq}}\delta t$ followed by a relaxation phase of length $t_{\textrm{run}}=N_{\textrm{run}}\delta t$, with $t_{\textrm{eq}}=1000/\omega_p$ and $t_{\textrm{run}}=300/\omega_p$.
The initial particle positions at time $t=-t_{\textrm{eq}}$ were assigned randomly in the simulation box, with a small region surrounding each particle excluded to avoid initial explosion.
The initial particle velocities ${\bf v}_i$ were sampled from a Maxwell-Boltzmann distribution at the desired temperature $T$, i.e. at the desired value of $\Gamma$.
During the equilibration phase, velocity scaling \cite{FrenkelSmit} was used to maintain the desired $\Gamma$ value.
At $t=0$, velocity scaling was turned off and the particle velocities were rescaled to the desired initial parallel and perpendicular temperatures as follows
\begin{eqnarray}
v_{i,x}(t=0^+)=v_{i,x}(0^-)\sqrt{\frac{T_\perp}{T}}\,,\,
v_{i,y}(0^+)=v_{i,y}(0^-)\sqrt{\frac{T_\perp}{T}}\,,\,
v_{i,z}(0^+)=v_{i,z}(0^-)\sqrt{\frac{T_\parallel}{T}} .
\end{eqnarray}
The system was then left to evolve freely, i.e. in the microcanonical ensemble, for a duration $t_{\textrm{run}}$.
During this period the instantaneous parallel and perpendicular temperatures, defined as follows in terms of the particles kinetic energies
\begin{eqnarray}
T_\parallel(t)\equiv\frac{2}{Nk_B}\sum_{i=1}^N{\frac{1}{2}mv_{i,z}^2(t)}\quad,\quad
T_\perp(t)\equiv\frac{1}{Nk_B}\sum_{i=1}^N{\frac{1}{2}m\left(v_{i,x}^2(t)+v_{i,y}^2(t)\right)},
\end{eqnarray}
were monitored.
In order to sample the initial statistical distribution function, Eq.~(\ref{eq:amax}), and to compare the MD simulations to the predictions of kinetic theories, $N_{\textrm{sim}}=25$ independent runs were performed with different initial conditions and averaged.
As illustrated in Fig.~\ref{fg:md_analysis}, the averaging smooths out the fluctuations in the kinetic energies inherent to each microcanonical dynamics.
 
To obtain the relaxation rate $\nu$, we assume that the MD temperatures evolve according to the rate equations (\ref{eq:nrl}).
The latter implies $\frac{d}{dt}\Delta T=-3\nu \Delta T$ with $\Delta T=T_\parallel-T_\perp$.
In general, as predicted by kinetic theory, $\nu$ depends on the time $t$ through its dependence upon  $T_\parallel$ and $T_\perp$.
If one assumes that the dependence is weak enough that $\nu$ is a constant on a short enough time scale $\Delta t$ beyond an initial time $t_0$ , then $\Delta T(t)=\Delta T(t_0)e^{-3\nu (t-t_0)}$ on this time scale (more sophisticated treatments that take into account the temperature dependence of $\nu$ are possible (e.g. \cite{dimo:08}) but we find that they do not affect the results in any significant manner).
The MD simulations confirm that $T_\parallel-T_\perp$ indeed decays exponentially for coupling parameters $\Gamma\!<\!1$ but only following a short transient period of duration $t_0$ of order $\omega_{p}^{-1}$; see Fig.~\ref{fg:md_analysis}b. For $\Gamma\geq 1$, as discussed in the next section, the temperature evolution is never exponential and the notion of relaxation rate defined based on the rate equations (\ref{eq:nrl}) does not apply.
The initial transient period describes the dependence on initial correlations, which are discarded in the kinetic theories discussed before.
In practice, the relaxation rates $\nu$ are obtained by fitting the MD data to the analytical solution $\Delta T(t_0)e^{-3\nu (t-t_0)}$ over the time interval $[t_0,t_0+\Delta t]$ with the time $t_0$ chosen right after the early transient behavior (the data shown below were obtained with $t_0=1/\omega_p$) and with $t_0+\Delta t$ chosen where the exponential behavior switches slope as a consequence of the dependence of $\nu$ on the time $t$. Figure~\ref{fg:md_analysis}b illustrates this procedure.

\section{Results \label{sec:results}}


\begin{figure}[t]
\sidecaption
\includegraphics[width=95mm]{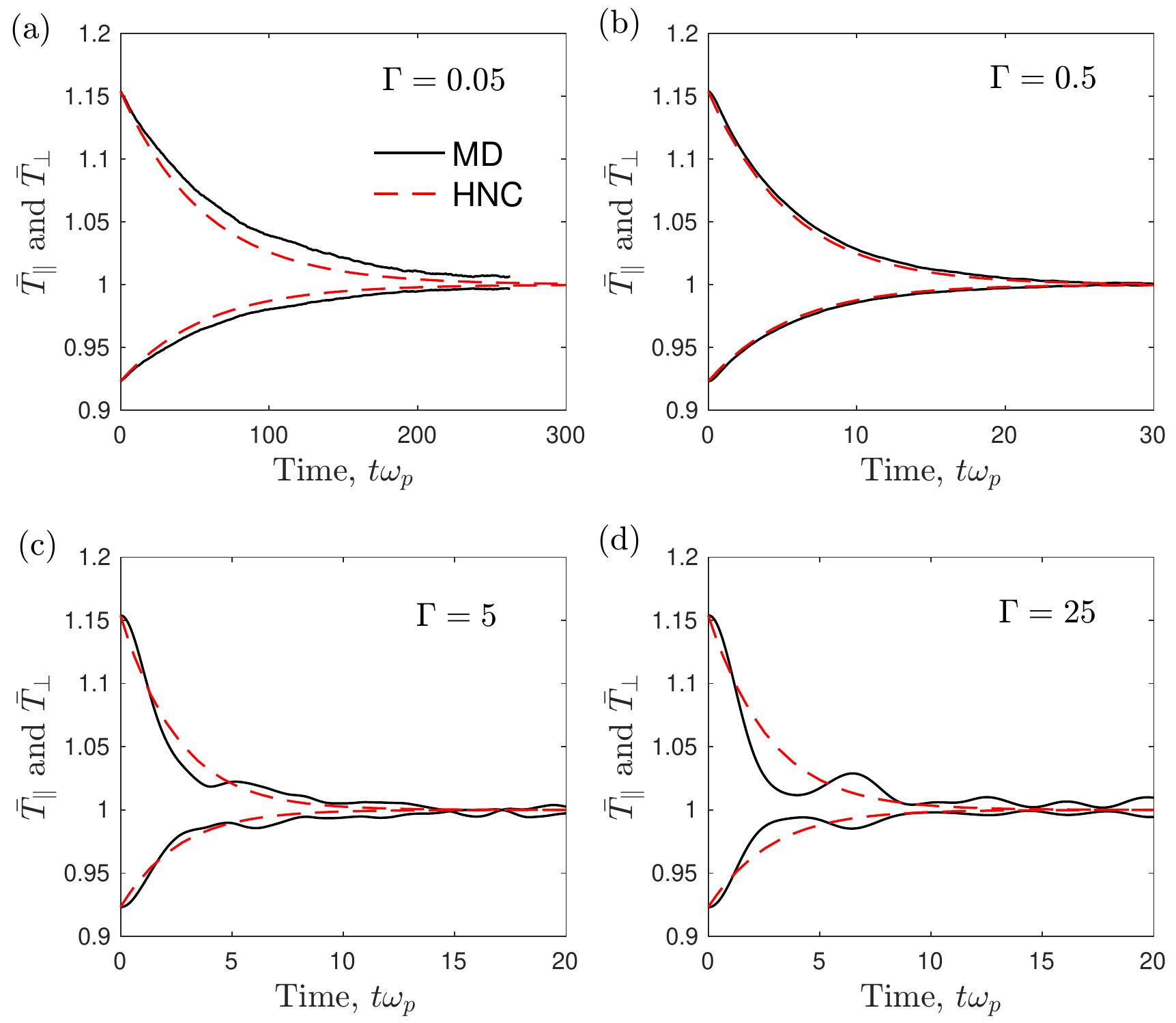}
\caption{Parallel and perpendicular temperature profiles in time computed from MD simulations (solid lines) at $\Gamma$ = 0.05, 0.5, 5 and 25 and initial temperatures of $\bar{T}_\parallel = T_\parallel/T = 1.15$ and $\bar{T}_\perp = T_\perp/T = 0.923$ ($A=-0.2$). Dashed lines show profiles predicted from the EPT approximation.  }
\label{fg:md_profiles}
\end{figure}

Figure~\ref{fg:md_profiles} shows parallel and perpendicular temperature profiles at four values of $\Gamma$ from MD simulations that were initiated with temperatures of $T_\parallel/T = 1.15$ and $T_\perp/T = 0.923$ ($A=-0.2$). Also shown are the profiles predicted by the EPT theory of Eq.~(\ref{eq:nu_ept}). The comparison shows good agreement with the predicted monotonically decreasing profiles at the lower $\Gamma$ values. 
As the coupling strength increases the rate of the relaxation remains well modeled, but features emerge in the MD data that are not predicted by the model. One feature is that the profiles from MD become non-monotonic, exhibiting oscillations. Such oscillations are common in time correlation functions at strong coupling. For instance, they can be observed in the velocity autocorrelation function, which when integrated provides the macroscopic diffusion coefficient~\cite{dali:06}. 
Dubin~\cite{dubi:05} has previously discussed how the temperature anisotropy relaxation rate can also be considered as the time integral of an energy correlation function. 

Another interesting feature of the MD data at strong coupling is a short initial delay, compared to the theoretical prediction, before relaxation onsets. This is also common feature of other correlation functions at strong coupling and is associated with initial conditions. For instance, the velocity autocorrelation function at short times behaves as $Z(t)/Z(0) = 1 - \Omega_E^2 t^2$, where $\Omega_E$ is the Einstein frequency~\cite{hans:06}. For an OCP, $\Omega_E = \omega_p/\sqrt{3}$. At weak coupling the relaxation time is much longer than $\omega_p^{-1}$, so this initial stage is so short that it is inconsequential in the overall relaxation profile. For instance, Fig.~\ref{fg:md_profiles}a shows that the relaxation time is approximately 200$\omega_p^{-1}$ at $\Gamma = 0.05$. However, at strong coupling the relaxation rate is only a few $\omega_p^{-1}$, so the short time rolloff on the $\Omega_E^{-1}$ timescale is a more pronounced feature of the overall relaxation profile. The data shown in Fig.~\ref{fg:md_profiles} suggests that similar arguments can also be applied to the temperature anisotropy relaxation process. All of the theories discussed above are Markovian, and as a consequence do not include information about these initial conditions. They all predict exponential profiles at short times, which leads to a linear-in-time term at the relaxation onset rather than a quadratic one. 

\begin{figure}[t]
\sidecaption
\includegraphics[width=75mm]{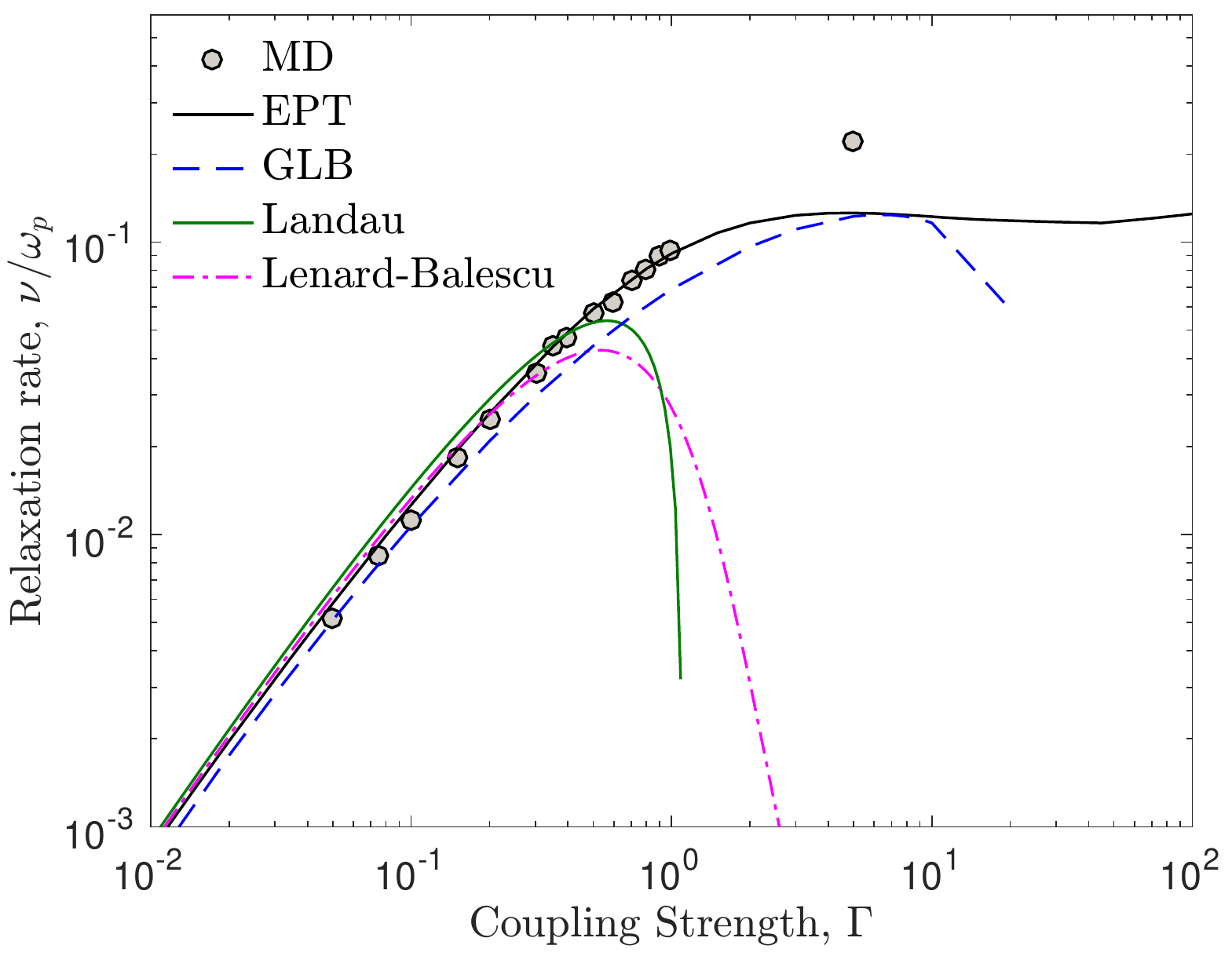}
\caption{Temperature anisotropy relaxation rate obtained from MD simulations (circles), and the predictions of the Landau theory from Eq.~(\ref{eq:nu_nrl}) (green line), the Lenard-Balescu theory from Eq.~(\ref{eq:nu_b0_full}) (magenta dash-dotted line), the EPT theory from Eq.~(\ref{eq:nu_ept}) (black line) and the generalized Lenard-Balescu theory from Eq.~(\ref{eq:glb_mid2}) (blue dashed line). }
\label{fg:md_compare}
\end{figure}

Figure~\ref{fg:md_compare} shows the results of the relaxation rate obtained from the MD data using the fitting procedure described in Sec.~\ref{sec:md}. The initial temperature anisotropy for this data set was the same $A=-0.2$ as from the data shown in Fig.~\ref{fg:md_profiles}. Predictions of the four theories described in Sections~\ref{sec:wc} and \ref{sec:sc} are shown for comparison. As expected, the weakly coupled theories compare well with the MD data at sufficiently small coupling strength ($\Gamma \lesssim 0.2$). As described in Sec.~\ref{sec:wc}, there is apparently no significant advantage to accounting for dynamic screening at these conditions. Both strongly coupled theories extend the standard plasma theories into the strongly coupled regime, but the generalized Lenard-Balescu theory from Sec.~\ref{sec:glb} predicts a negative collision frequency for $\Gamma \gtrsim 20$ (negative values are not shown on this log-log scale). This feature of the theory has also been noted in computing the viscosity coefficient in \cite{dali:14}. Although MD simulations were carried out at higher $\Gamma$ values than shown in this figure, the oscillatory nature of the temperature profiles made the fitting procedure unreasonable at sufficiently high coupling strength.

Figures~\ref{fg:md_profiles}a and \ref{fg:md_compare} show that for $\Gamma \lesssim 0.1$ the relaxation rate predicted by all of the theories is slightly larger than what is observed in the MD simulations (for example, by approximately 20\% at $\Gamma = 0.05$). This is likely associated with the assumption made in the theoretical analysis that the distributions maintain the anisotropic Maxwellian form. It is well known in other transport processes that distortions of the distribution function contributes an order unity correction to the transport rates at weak coupling. This is accounted for in hydrodynamic theories, such as Chapman-Enskog or Grad, through the higher order terms of the Sonine polynomial expansions of the distribution function. However, the magnitude of these terms is known to rapidly diminish as $\Gamma \gtrsim 1$. For instance, it was shown in \cite{baal:14} that for self-diffusion of the OCP the approximately 20\% contribution of the second order term of the expansion vanishes for $\Gamma > 1$ (see figure 9 of that reference). Figure~\ref{fg:md_compare} shows evidence for the same effect in the temperature anisotropy relaxation rate. To the best of our knowledge, such an expansion procedure has not yet been applied to temperature anisotropy relaxation.

\begin{figure}[t]
\sidecaption
\includegraphics[width=160mm]{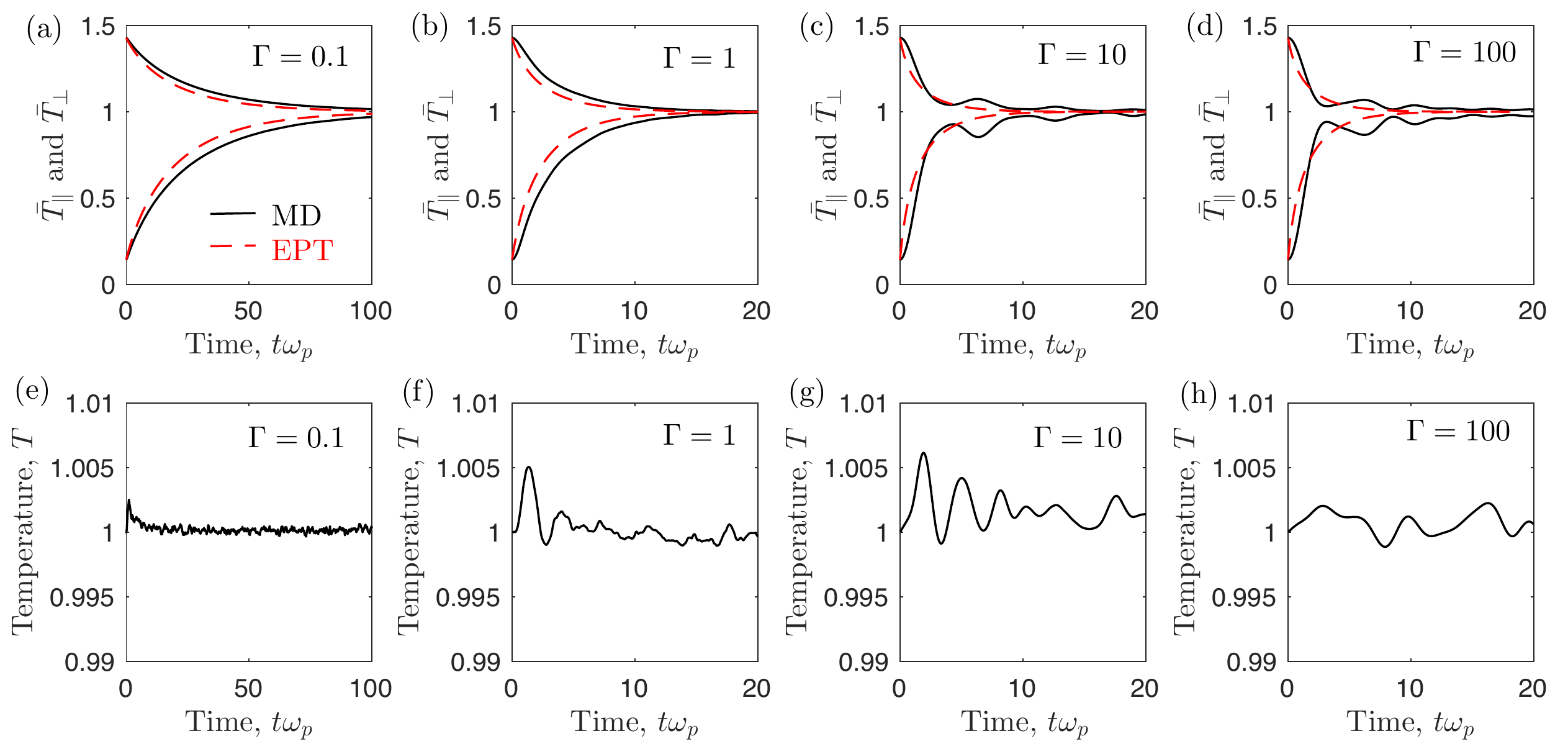}
\caption{Time dependent temperature profiles with a highly anisotropic initial condition of $T_\perp/T_\parallel = 10$ ($A=9$). Panels (a)--(d) show the parallel ($\bar{T}_\parallel = T_\parallel/T$) and perpendicular ($\bar{T}_\perp = T_\perp/T$) temperatures from MD simulations (black solid lines) and predictions from EPT (red dashed lines). Panels (e)--(h) show the total kinetic temperature as a function of time in terms of its initial value ($T(t)/T_o$).}
\label{fg:large_a_profiles}
\end{figure}

Finally, Fig.~\ref{fg:large_a_profiles} shows temperature relaxation profiles from MD simulations at $\Gamma = 0.1$, 1, 10 and 100 conducted with a large initial temperature anisotropy of $T_\perp/T_\parallel = 10$ ($A=9$). The comparison with the EPT predictions is similar to what was observed at smaller initial anisotropy. The accuracy is, perhaps, slightly less than was observed from the small anisotropy cases in Fig.~\ref{fg:md_profiles} near $\Gamma \simeq 1$, but the generally good agreement over this entire range of coupling strength provides strong evidence that the theory is robust even at a very large initial anisotropy. The predicted rate compares well even at $\Gamma = 100$. Dynamic screening apparently is not significant even at $T_\perp/T_\parallel =10$. Panels e-h show the total temperature as a function of time for the same simulations. These show that at weak coupling the total temperature, and thus total kinetic energy, does not fluctuate more than approximately 0.1\% throughout the evolution of the system. The higher $\Gamma$ values show that the temperature fluctuates in time by as much as 0.5\%. Recall that each curve is obtained from 25 independent simulations.  The total energy of the system was confirmed to be conserved to better than 1 part in $10^6$ throughout each simulation. These oscillations are not numerical artifacts, but rather represent oscillations in the exchange between kinetic and potential energy associated with correlations as the system relaxes. None of the theories discussed address potential energy of the system, and thus do not model this effect. 

\section{Conclusions} 

This comparison between two common approaches to plasma kinetic theory and MD simulations has suggested a few general conclusions with regard to the temperature anisotropy relaxation, and has also revealed a few remaining gaps in current understanding. The data comparison suggests that dynamic screening, which is modeled in the Lenard-Balescu and generalized Lenard-Balescu theories, does not significantly influence the relaxation rate. The Boltzmann-based approaches (Landau or EPT) can be accurately applied in most situations of practical interest, rather than Lenard-Balescu based theories which are comparatively difficult to evaluate. Comparison with MD revealed that the EPT provides an accurate approach to modeling the anisotropy relaxation rate over a similar range of coupling strength as has been encountered for other processes, such as diffusion~\cite{baal:15}. However, none of the theories capture kinetic energy oscillations (associated with kinetic-potential energy exchange) or early time delay (associated with initial conditions) in the time-dependent temperature profiles that were observed in the MD simulations. These effects are associated with strong correlations. Finally, one additional limitation of current models is the assumption that the distribution maintains the specified anisotropic Maxwellian form throughout the evolution.  Comparison with MD reveals that improvements to predicted relaxation rates may be expected from accounting for deviations from this assumed distribution. 

\begin{acknowledgement}
  This material is based upon work supported by LDRD project 20150520ER at Los Alamos National Laboratory. The work of SDB was also supported by the Air Force Office of Scientific Research under award number FA9550-16-1-0221 and by the National Science Foundation under Grant No. PHY-1453736.

\end{acknowledgement}

%
%

\end{document}